\documentstyle[sprocl,epsfig]{article}

\def\Journal#1#2#3#4{{#1} {\bf #2}, #3 (#4)}

\def\NPB{{\em Nucl.\ Phys.} B}
\def\PLB{{\em Phys.\ Lett.}  B}

\def\ZPC{{\em Z.\ Phys.} C}

\def\be{\begin{equation}}
\def\ee{\end{equation}}
\def\bea{\begin{eqnarray}}
\def\eea{\end{eqnarray}}

\def\kf{{\bf k}}
\def\qf{{\bf q}}
\def\lf{{\bf l}}

\begin{document}
\renewcommand{\thefootnote}{\fnsymbol{footnote}}

\hfill DESY 97-130

\vspace{1cm}

\title{TOWARDS AN EFFECTIVE THEORY OF SMALL-X QCD 
\footnote{Talk presented at the International School of 
Subnuclear Physics, 34th Course: {\sl Effective Theories and Fundamental 
Interactions}, Erice, Sicily, 3--12 July 1996; to appear in the proceedings.}
}

\author{ C.\ EWERZ }

\address{II.\ Institut f\"ur Theoretische Physik, Universit\"at Hamburg,\\
Hamburg, Germany}




\maketitle\abstracts{
In this talk I describe recent progress in investigating 
the high energy limit of perturbative QCD. 
I review some of the steps that have been done in the direction 
of constructing an effective field theory for this limit. 
I describe some of its building blocks and 
explain why we expect the effective theory to be a $2+1$--dimensional
conformal field theory.
}
  
\section{Introduction}
In this talk I would like to describe some recent developments 
in the field of small--$x$ perturbative QCD. In particular, I will  
review the first steps that have been done in the direction 
of finding an effective theory for small--$x$ QCD. 

Regge physics has been around for thirty years now. 
The focus of Regge physics is the behavior 
of hadronic scattering amplitudes at very high 
energy $s$ and fixed momentum transfer $t$ (of the order of some hadronic 
mass scale),  
\be
  s \gg t \simeq M^2_{\mbox{\scriptsize hadron}}  \,.
\ee
With the advent of the HERA machine, it has recently attracted new 
interest. Electron--proton collisions in the now available 
kinematic range of small Bjorken--$x$ and large momentum transfer $Q^2$ 
allow for a new test of high energy QCD. 
The interesting subprocess here is the scattering of a highly 
virtual photon off the proton. 
From a theorists point of view it is particularly interesting, 
because the large photon virtuality $Q^2$ enables us to enter this region 
of high parton densities using perturbative methods. 

Let us first consider hadron--hadron scattering. The optical 
theorem relates the total cross section 
to the imaginary part of the forward elastic scattering amplitude, 
\be
  \sigma_{\mbox{\scriptsize tot}} = 
  \frac{1}{s}\, \mbox{Im}\, A_{\mbox{\scriptsize el}}(s,t=0) \,.
\label{opttheo}
\ee
It is convenient to perform a Mellin transformation 
changing from energy $s$ to complex angular momentum $\omega$,
\be
  A(s,t) = i s \int_{\delta-i \infty}^{\delta+i \infty} 
           \frac{d\omega}{2 \pi i} \left( \frac{s}{M^2} \right)^\omega
           A(\omega,t) \,.
\label{Mellin}
\ee
The high energy behavior of the total cross section is 
then determined by the singularities of $A(\omega,t)$ in 
the $\omega$--plane, the so--called Regge poles and Regge cuts. 
The rightmost singularity gives the leading 
contribution and is identified with the pomeron. 
As it describes an elastic amplitude (see (\ref{opttheo})) it
carries vacuum quantum numbers. 
The location of the Regge poles in hadron--hadron scattering 
cannot yet be calculated from first principles, but 
there are successful phenomeological models 
that describe the experimental data in this framework. 

It follows from unitarity that the total cross section 
cannot grow infinitely fast at high energy. It has to 
satisfy the Froissart bound
\be
  \sigma_{\mbox{\scriptsize tot}} \le \mbox{const.} \,\log^2(s) \,.
\label{Froissart}
\ee

In the remaining part of my talk I will concentrate 
on considerations applying to the processes of 
$\gamma^\ast p$ or $\gamma^\ast \gamma^\ast$ scattering. 
The high virtuality $Q^2$ of the photon allows a 
perturbative treatment and thus enables us to get a clear 
understanding of these processes. 
$\gamma^\ast$--proton scattering 
is the basic process in electron--proton collisions. 
The Bjorken--$x$ variable is at high energy given 
by $x \simeq \frac{Q^2}{s}$, high energy corresponds to 
small $x$. 

At small $x$ the small value of the coupling constant $\alpha_s$ 
can be compensated by large logarithms of $x$. 
This leads us to the leading logarithmic approximation (LLA) 
\be
  \alpha_s \ll 1 \,;\:\:\: \alpha_s \log(1/x) \sim 1 \,.
\ee
In this approximation, the infinite number of 
contributing diagrams can be resummed. The result is the BFKL 
equation.\cite{BFKL} It describes the $t$--channel exchange of 
a bound state of two reggeized gluons. These reggeized gluons 
are collective excitations of the Yang--Mills field carrying gluon 
quantum numbers. They are the relevant degrees of freedom at small $x$. 

The BFKL equation is an integral equation in two--dimensional transverse 
momentum space, because the longitudinal degrees of freedom 
decouple in the high energy limit. In detail, it has the form 
\be
  \omega \phi_\omega(\kf,\qf-\kf) = \phi^0(\kf,\qf-\kf) 
              + \int \frac{d^2\kf'}{(2\pi)^3} \,
              \frac{1}{\kf'^2 (\qf-\kf')^2} 
              {\cal K}(\qf,\kf,\kf') \,\phi_\omega(\kf',\qf-\kf') \,.
\label{BFKLeq}
\ee
$\kf$, $\qf-\kf$ are the momenta of the two gluons, $\phi^0$ is an 
inhomogeneous term. 
The integral kernel ${\cal K}$ (the so--called Lipatov kernel) is given 
by
\bea
{\cal K}(\kf,\kf',\qf) &=& 
        N_c g^2 \left[ -\qf^2 + \frac{\kf^2(\qf-\kf')^2}{(\kf-\kf')^2} 
            + \frac{(\qf-\kf)^2 \kf'^2}{(\kf-\kf')^2} \right] \nonumber \\
   & & - (2\pi)^3 \kf^2 (\qf-\kf)^2 
        \left[ \beta(\kf) + \beta(\qf - \kf) \right] 
        \delta^{(2)}(\kf-\kf')
  \,.
\eea
The coupling constant is normalized to $\alpha_s=\frac{g^2}{4\pi}$ and 
$\alpha(\lf^2) = 1 + \beta(\lf^2)$ with 
\be
  \beta(\lf^2) = \frac{N_c}{2} g^2  \int \frac{d^2\kf'}{(2 \pi)^3} 
          \frac{\lf^2}{\kf'^2 (\kf' -\lf)^2} \,.
\ee
is known as the gluon trajectory 
function. 
The variable $\omega$ acts as an energy variable in the BFKL 
equation. It can be shown to be conjugate to rapidity. 
Without making use of the Mellin transformation 
we would have found the BFKL equation as an evolution equation in $x$. 

The general form of the BFKL amplitude can be derived 
from the integral equation by iteration. 
Thus the high energy elastic scattering amplitude is in LLA 
described by a $t$--channel exchange of gluon ladders: 
\be
  \begin{picture}(0,0)%
\epsfig{file=ladder.pstex}%
\end{picture}%
\setlength{\unitlength}{0.00083300in}%
\begingroup\makeatletter\ifx\SetFigFont\undefined
\def\x#1#2#3#4#5#6#7\relax{\def\x{#1#2#3#4#5#6}}%
\expandafter\x\fmtname xxxxxx\relax \def\y{splain}%
\ifx\x\y   
\gdef\SetFigFont#1#2#3{%
  \ifnum #1<17\tiny\else \ifnum #1<20\small\else
  \ifnum #1<24\normalsize\else \ifnum #1<29\large\else
  \ifnum #1<34\Large\else \ifnum #1<41\LARGE\else
     \huge\fi\fi\fi\fi\fi\fi
  \csname #3\endcsname}%
\else
\gdef\SetFigFont#1#2#3{\begingroup
  \count@#1\relax \ifnum 25<\count@\count@25\fi
  \def\x{\endgroup\@setsize\SetFigFont{#2pt}}%
  \expandafter\x
    \csname \romannumeral\the\count@ pt\expandafter\endcsname
    \csname @\romannumeral\the\count@ pt\endcsname
  \csname #3\endcsname}%
\fi
\fi\endgroup
\begin{picture}(3087,980)(901,-1715)
\put(2851,-1261){\makebox(0,0)[lb]{\smash{\SetFigFont{10}{12.0}{rm}${\huge \sum}$}}}
\put(2326,-1261){\makebox(0,0)[lb]{\smash{\SetFigFont{10}{12.0}{rm}$=$}}}
\put(2776,-1486){\makebox(0,0)[lb]{\smash{\SetFigFont{10}{12.0}{rm}{\scriptsize of rungs}}}}
\put(2776,-1411){\makebox(0,0)[lb]{\smash{\SetFigFont{10}{12.0}{rm}{\scriptsize number}}}}
\put(901,-1261){\makebox(0,0)[lb]{\smash{\SetFigFont{10}{12.0}{rm}$s \to \infty$}}}
\end{picture}
 \,,
\ee
the rungs being Lipatov kernels. 
The BFKL equation has been solved analytically and leads to 
an increase of the amplitude 
\be
  A \sim x^{-(1+\omega_{\mbox{\tiny BFKL}})}\,; \:\:\:
  \omega_{{\mbox{\scriptsize BFKL}}} = \frac{\alpha_s N_c}{\pi} \,4 \ln 2 
   \simeq 0.5  
\ee
for $t=0$ in the limit $x \to 0$. 
It follows that the Froissart bound is violated 
at very small $x$, 
\be
  \sigma_{\mbox{\scriptsize tot}}^{\gamma^\ast p}
  \sim x^{-\omega_{\mbox{\tiny BFKL}}}
  \not\le \mbox{const.} \,\log^2(1/x) \,.
\ee
This means that the BFKL pomeron violates unitarity at very small $x$. 

\section{Unitarization of the BFKL-Pomeron}
To restore unitarity we have to include nonleading (in $\log(1/x)$) 
corrections to the BFKL equation. 
The minimal set of corrections restoring unitarity can 
be identified as the corrections with a 
larger number of reggeized gluons in the 
$t$--channel.\cite{moregluons} 
Our goal is to find a consistent framework for the description of the 
infinite number of these terms. 
This framework will be a more general structure in which the 
BFKL pomeron appears as the first approximation. 

Based on our knowledge of the BFKL equation 
we can already state some properties of this effective theory. 
Like in the BFKL case, the longitudinal degrees of freedom 
can be factored off and 
the dynamics will take place in two--dimensional transverse momentum space. 
Again, rapidity will act as the time--like coordinate. 
The appropriate degrees of freedom will be reggeized gluons. 

As I will show, very important new elements are number changing vertices 
describing the creation and annihilation of $t$--channel gluons.
Thus our aim to construct an effective theory of QCD at small $x$ 
turns out to be that of building a $2+1$--dimensional 
quantum field theory of reggeized gluons. 

\begin{figure}
  \begin{center}
    \leavevmode
     \input{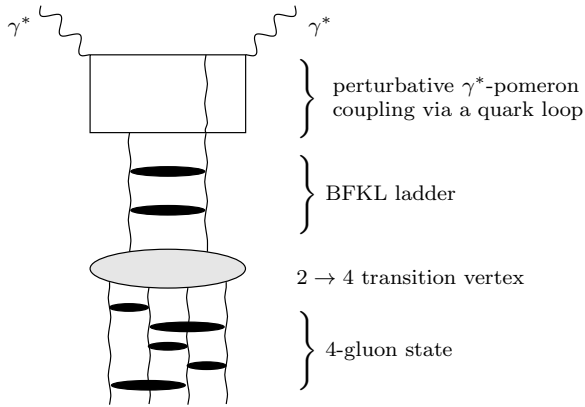}
  \end{center}
\caption{The basic elements of the $2+1$--dimensional 
field theory for small $x$ QCD
\label{fig:structure}}
\end{figure}

The program to be carried out in order to achieve this ambitious goal 
can be cut into the following pieces. Figure \ref{fig:structure} tries 
to visualize the different elements that have to be extracted from QCD.

\begin{itemize}
\item {\em The leading order elements} 
are known: the BFKL equation and the quark loop.  

\item {\em The next--to--leading corrections.} 
The NLLA corrections to the BFKL equation are needed, for example, to 
understand which symmetries of the BFKL equation are only 
due to the LLA and how their breaking occurs. 
These corrections are currently being calculated 
by Lipatov and Fadin\,\cite{NLLA} and by 
Ciafaloni and Camici.\cite{Ciafaloni}

\item {\em The spectrum of the $n$--gluon state} 
is a quantum mechanical problem formally described by the 
so--called BKP equations.\cite{moregluons,BKP}
Recently there has been a great leap forward in this longstanding 
problem. Lipatov \cite{LipXXX}, Faddeev and Kor\-chemsky \cite{XXX} 
were able to prove that it is, in the large $N_c$ limit, 
equivalent to the integrable XXX Heisenberg 
model for noncompact spin zero. 
This connection opens the field for the  
application of some very powerful mathematical methods (Bethe ansatz etc.). 

\item {\em Number changing vertices,} 
which turn the problem into that of a quantum field theory. 
The vertex $2 \rightarrow 4$ gluons has been found recently,\cite{BarWust} 
higher transition vertices, especially the $2\rightarrow 6$ transition, 
are currently under investigation.\cite{d6} 

\item {\em Finding symmetries} of the above elements is of course 
obligatory to understand the structure of the field theory.
The most important symmetry of the known elements is their 
conformal invariance.
\end{itemize}
The number changing vertices, their emergence and the conformal 
invariance will be the topic of the following sections. 

\section{The Number Changing Vertices}
The formal framework for our considerations is made up by 
amplitudes $D_n$ describing the creation of $n$ reggeized gluons 
in the $t$--channel from a quark loop at fixed $\omega$ 
(see fig.\ \ref{fig:structure}). 
These amplitudes will depend on the two--dimensional transverse 
momenta and on the color indices of the gluons.
The amplitudes obey a tower of coupled 
integral equations that generalize the 
BFKL equation. For the purpose of this talk we 
state the equations only in a graphical form and only up to 
$n=4$, a more detailed account 
can be found elsewhere.\cite{BarWust,d6}
\bea
  & & \begin{picture}(0,0)%
\epsfig{file=gld2.pstex}%
\end{picture}%
\setlength{\unitlength}{0.00083300in}%
\begingroup\makeatletter\ifx\SetFigFont\undefined
\def\x#1#2#3#4#5#6#7\relax{\def\x{#1#2#3#4#5#6}}%
\expandafter\x\fmtname xxxxxx\relax \def\y{splain}%
\ifx\x\y   
\gdef\SetFigFont#1#2#3{%
  \ifnum #1<17\tiny\else \ifnum #1<20\small\else
  \ifnum #1<24\normalsize\else \ifnum #1<29\large\else
  \ifnum #1<34\Large\else \ifnum #1<41\LARGE\else
     \huge\fi\fi\fi\fi\fi\fi
  \csname #3\endcsname}%
\else
\gdef\SetFigFont#1#2#3{\begingroup
  \count@#1\relax \ifnum 25<\count@\count@25\fi
  \def\x{\endgroup\@setsize\SetFigFont{#2pt}}%
  \expandafter\x
    \csname \romannumeral\the\count@ pt\expandafter\endcsname
    \csname @\romannumeral\the\count@ pt\endcsname
  \csname #3\endcsname}%
\fi
\fi\endgroup
\begin{picture}(2927,548)(901,-373)
\put(2701,-61){\makebox(0,0)[lb]{\smash{\SetFigFont{10}{12.0}{rm}$+$}}}
\put(1726,-61){\makebox(0,0)[lb]{\smash{\SetFigFont{10}{12.0}{rm}$=$}}}
\put(901,-61){\makebox(0,0)[lb]{\smash{\SetFigFont{10}{12.0}{rm}$\omega$}}}
\put(3152,-62){\makebox(0,0)[lb]{\smash{\SetFigFont{8}{9.6}{rm}$D_2$}}}
\put(2088,-47){\makebox(0,0)[lb]{\smash{\SetFigFont{8}{9.6}{rm}$D_{(2;0)}$}}}
\put(1276,-62){\makebox(0,0)[lb]{\smash{\SetFigFont{8}{9.6}{rm}$D_2$}}}
\end{picture}
 \\
  & & \begin{picture}(0,0)%
\epsfig{file=gld3.pstex}%
\end{picture}%
\setlength{\unitlength}{0.00083300in}%
\begingroup\makeatletter\ifx\SetFigFont\undefined
\def\x#1#2#3#4#5#6#7\relax{\def\x{#1#2#3#4#5#6}}%
\expandafter\x\fmtname xxxxxx\relax \def\y{splain}%
\ifx\x\y   
\gdef\SetFigFont#1#2#3{%
  \ifnum #1<17\tiny\else \ifnum #1<20\small\else
  \ifnum #1<24\normalsize\else \ifnum #1<29\large\else
  \ifnum #1<34\Large\else \ifnum #1<41\LARGE\else
     \huge\fi\fi\fi\fi\fi\fi
  \csname #3\endcsname}%
\else
\gdef\SetFigFont#1#2#3{\begingroup
  \count@#1\relax \ifnum 25<\count@\count@25\fi
  \def\x{\endgroup\@setsize\SetFigFont{#2pt}}%
  \expandafter\x
    \csname \romannumeral\the\count@ pt\expandafter\endcsname
    \csname @\romannumeral\the\count@ pt\endcsname
  \csname #3\endcsname}%
\fi
\fi\endgroup
\begin{picture}(4052,571)(601,-396)
\put(2852,-62){\makebox(0,0)[lb]{\smash{\SetFigFont{8}{9.6}{rm}$D_2$}}}
\put(601,-61){\makebox(0,0)[lb]{\smash{\SetFigFont{10}{12.0}{rm}$\omega$}}}
\put(1426,-61){\makebox(0,0)[lb]{\smash{\SetFigFont{10}{12.0}{rm}$=$}}}
\put(2401,-61){\makebox(0,0)[lb]{\smash{\SetFigFont{10}{12.0}{rm}$+$}}}
\put(3301,-61){\makebox(0,0)[lb]{\smash{\SetFigFont{10}{12.0}{rm}$+\,\,\, \sum$}}}
\put(3977,-62){\makebox(0,0)[lb]{\smash{\SetFigFont{8}{9.6}{rm}$D_3$}}}
\put(977,-62){\makebox(0,0)[lb]{\smash{\SetFigFont{8}{9.6}{rm}$D_3$}}}
\put(1787,-43){\makebox(0,0)[lb]{\smash{\SetFigFont{8}{9.6}{rm}$D_{(3;0)}$}}}
\end{picture}
 \label{d3eq} \\
  & & \begin{picture}(0,0)%
\epsfig{file=gld4.pstex}%
\end{picture}%
\setlength{\unitlength}{0.00083300in}%
\begingroup\makeatletter\ifx\SetFigFont\undefined
\def\x#1#2#3#4#5#6#7\relax{\def\x{#1#2#3#4#5#6}}%
\expandafter\x\fmtname xxxxxx\relax \def\y{splain}%
\ifx\x\y   
\gdef\SetFigFont#1#2#3{%
  \ifnum #1<17\tiny\else \ifnum #1<20\small\else
  \ifnum #1<24\normalsize\else \ifnum #1<29\large\else
  \ifnum #1<34\Large\else \ifnum #1<41\LARGE\else
     \huge\fi\fi\fi\fi\fi\fi
  \csname #3\endcsname}%
\else
\gdef\SetFigFont#1#2#3{\begingroup
  \count@#1\relax \ifnum 25<\count@\count@25\fi
  \def\x{\endgroup\@setsize\SetFigFont{#2pt}}%
  \expandafter\x
    \csname \romannumeral\the\count@ pt\expandafter\endcsname
    \csname @\romannumeral\the\count@ pt\endcsname
  \csname #3\endcsname}%
\fi
\fi\endgroup
\begin{picture}(5177,544)(1501,-1570)
\put(4877,-1262){\makebox(0,0)[lb]{\smash{\SetFigFont{8}{9.6}{rm}$D_3$}}}
\put(5326,-1261){\makebox(0,0)[lb]{\smash{\SetFigFont{10}{12.0}{rm}$+\,\,\, \sum$}}}
\put(1501,-1261){\makebox(0,0)[lb]{\smash{\SetFigFont{10}{12.0}{rm}$\omega$}}}
\put(2326,-1261){\makebox(0,0)[lb]{\smash{\SetFigFont{10}{12.0}{rm}$=$}}}
\put(3301,-1261){\makebox(0,0)[lb]{\smash{\SetFigFont{10}{12.0}{rm}$+$}}}
\put(4201,-1261){\makebox(0,0)[lb]{\smash{\SetFigFont{10}{12.0}{rm}$+\,\,\, \sum$}}}
\put(6002,-1262){\makebox(0,0)[lb]{\smash{\SetFigFont{8}{9.6}{rm}$D_4$}}}
\put(3752,-1262){\makebox(0,0)[lb]{\smash{\SetFigFont{8}{9.6}{rm}$D_2$}}}
\put(1877,-1262){\makebox(0,0)[lb]{\smash{\SetFigFont{8}{9.6}{rm}$D_4$}}}
\put(2691,-1248){\makebox(0,0)[lb]{\smash{\SetFigFont{8}{9.6}{rm}$D_{(4;0)}$}}}
\end{picture}
 \label{d4eq}
\eea
The inhomogeneous terms $D_{(n;0)}$ stand for the perturbative 
coupling of $n$ gluons directly to the light quark loop. 
The first equation is the usual BFKL equation. 
The kernels appearing in the other equations can be calculated 
perturbatively and generalize the Lipatov kernel. 
They must not be confused with the number changing 
vertices we want to calculate. The latter emerge when we 
solve the integral equations which we can do at least partially, 
as I will explain. 
The summation symbols indicate that we have to include 
all possible pairwise interactions of $t$--channel gluons. 

The equation for $D_3$ can be solved exactly and the solution is 
\bea
\lefteqn{D_3^{a_1a_2a_3}(\kf_1,\kf_2,\kf_3) 
= C_3 g f_{a_1a_2a_3} \times} \nonumber \\
& & \times (D_2(\kf_1+\kf_2,\kf_3) - D_2(\kf_1+\kf_3,\kf_2) 
+ D_2(\kf_1,\kf_2+\kf_3)) \,,
\label{d3d2}
\eea
where $a_i$ are color indices, $\kf_i$ are the transverse momenta 
of the gluons and $C_3$ is a normalization constant. 
This result tells us that, although $D_3$ is formally defined as 
a 3--gluon amplitude, it is --- according to its analytic properties 
(\ref{d3d2}) --- 
actually a superposition of 2--gluon (i.\,e.\ 
BFKL) amplitudes. There is no intermediate 3--gluon state! 
The  details of the calculation show that this fact strongly 
depends on the inhomogeneous term in the integral equation (\ref{d3eq}), 
namely the coupling of the gluons to the loop containing 
light (effectively massless) quarks. 

The equation (\ref{d4eq}) for the 4--gluon amplitude can be solved 
at least partially. 'Partially' because the solution involves the 
full 4--gluon state which is not yet known. Nevertheless, we can 
extract the structure of the solution from the equation. 
For simplicity suppressing all color and normalization factors, 
it has the following form: 
\be
  \begin{picture}(0,0)%
\epsfig{file=solutiond4.pstex}%
\end{picture}%
\setlength{\unitlength}{0.00083300in}%
\begingroup\makeatletter\ifx\SetFigFont\undefined
\def\x#1#2#3#4#5#6#7\relax{\def\x{#1#2#3#4#5#6}}%
\expandafter\x\fmtname xxxxxx\relax \def\y{splain}%
\ifx\x\y   
\gdef\SetFigFont#1#2#3{%
  \ifnum #1<17\tiny\else \ifnum #1<20\small\else
  \ifnum #1<24\normalsize\else \ifnum #1<29\large\else
  \ifnum #1<34\Large\else \ifnum #1<41\LARGE\else
     \huge\fi\fi\fi\fi\fi\fi
  \csname #3\endcsname}%
\else
\gdef\SetFigFont#1#2#3{\begingroup
  \count@#1\relax \ifnum 25<\count@\count@25\fi
  \def\x{\endgroup\@setsize\SetFigFont{#2pt}}%
  \expandafter\x
    \csname \romannumeral\the\count@ pt\expandafter\endcsname
    \csname @\romannumeral\the\count@ pt\endcsname
  \csname #3\endcsname}%
\fi
\fi\endgroup
\begin{picture}(3801,1325)(751,-2233)
\put(4427,-1861){\makebox(0,0)[lb]{\smash{\SetFigFont{10}{12.0}{rm}$V$}}}
\put(3376,-1561){\makebox(0,0)[lb]{\smash{\SetFigFont{10}{12.0}{rm}$+$}}}
\put(751,-1561){\makebox(0,0)[lb]{\smash{\SetFigFont{10}{12.0}{rm}$D_4(\kf_1,\kf_2,\kf_3,\kf_4) = \:\sum$}}}
\end{picture}
 \label{solutiond4}
\ee
The first term is the sum of 2--gluon (BFKL) amplitudes $D_2$. 
The summation extends over the (seven) possibilities to combine the four 
momenta into two sums which are then the two arguments of the $D_2$ 
amplitudes ($C_{1,2}$ indicate color tensors, but again we will not 
go into the details of $su(3)$ algebra.):
\bea
  D_4^R(\kf_1,\kf_2,\kf_3,\kf_4) &=& 
   C_1\, [ D_2(\kf_1+\kf_3+\kf_4,\kf_2) + D_2(\kf_1+\kf_2+\kf_4,\kf_3)
   \nonumber \\
 & &         - D_2(\kf_1+\kf_2,\kf_3+\kf_4) - D_2(\kf_1+\kf_3,\kf_2+\kf_4) ]
   \nonumber \\
 &+&    C_2\, [ D_2(\kf_1+\kf_2+\kf_3,\kf_4) + D_2(\kf_1,\kf_2+\kf_3+\kf_4) 
   \nonumber \\
 & &         - D_2(\kf_1+\kf_4,\kf_2+\kf_3) ]
\eea
This first part $D_4^R$ being isolated, the remaining 
terms in the equation define the transition 
vertex $V_{2 \rightarrow 4}$. (The full derivation of the vertex 
is rather lengthy and will not be given here.) 
We can write the second term as the convolution 
$G_4 \otimes V_{2 \rightarrow 4} \otimes D_2$ with
$G_4$ being the propagator of the 4--gluon state. 
Starting from the quark loop, we first have a 2--gluon state, then 
there is a transition from two to four reggeized gluons and at the bottom 
we have the full, i.\,e.\ interacting, 4--gluon state. 
This teaches us the important lesson 
that in leading logarithmic order 
it is not possible to couple the interacting system 
of four reggeized gluons directly to the quark loop! 
This coupling always involves the vertex 
$V_{2 \rightarrow 4}$ and the 2--gluon state. 

Let us now have a closer look at the vertex $V_{2 \rightarrow 4}$. 
It can be written as 
\bea
 V_{2 \rightarrow 4}^{a_1a_2a_3a_4}(\{ {\bf q}_j\},\kf_1,\kf_2,\kf_3,\kf_4) 
 &=& 
  \delta_{a_1a_2} \delta_{a_3a_4} V(\{ {\bf q}_j\},\kf_1,\kf_2;\kf_3,\kf_4)
 \nonumber \\
 & & 
 + \delta_{a_1a_3} \delta_{a_2a_4} V(\{ {\bf q}_j\},\kf_1,\kf_3;\kf_2,\kf_4)
 \nonumber \\
 & &
 + \delta_{a_1a_4} \delta_{a_2a_3} V(\{ {\bf q}_j\},\kf_1,\kf_4;\kf_2,\kf_3)
\label{colV}
\eea
Again, the indices $a_i$ are color labels, $\kf_i$ are the outgoing 
momenta and $\{ {\bf q}_j\}$ the incoming momenta. 
The explicit analytic form of the function $V$ can be found in the 
literature.\cite{BarWust} 
The representation (\ref{colV}) 
nicely demonstrates the complete symmetry 
of the vertex under permutations of the outgoing gluons and 
displays its simple color structure. 

Essentially in the same way, it is possible to extract information 
about the higher $n$--gluon amplitudes from the corresponding 
equations even without explicit knowledge of the interacting 
$n$--gluon systems. 

A new result\,\cite{d6} I can present here is the following: 
Like in the case of the 3--gluon amplitude, 
also the 5--gluon intermediate state is absent, 
the mechanism being similar to that shown in (\ref{d3d2}): 
the 5--gluon amplitude $D_5$ can be decomposed into a sum of 4--gluon 
amplitudes $D_4$. Using the same notation as in (\ref{solutiond4}), 
\be
  \begin{picture}(0,0)%
\epsfig{file=solutiond5.pstex}%
\end{picture}%
\setlength{\unitlength}{0.00083300in}%
\begingroup\makeatletter\ifx\SetFigFont\undefined
\def\x#1#2#3#4#5#6#7\relax{\def\x{#1#2#3#4#5#6}}%
\expandafter\x\fmtname xxxxxx\relax \def\y{splain}%
\ifx\x\y   
\gdef\SetFigFont#1#2#3{%
  \ifnum #1<17\tiny\else \ifnum #1<20\small\else
  \ifnum #1<24\normalsize\else \ifnum #1<29\large\else
  \ifnum #1<34\Large\else \ifnum #1<41\LARGE\else
     \huge\fi\fi\fi\fi\fi\fi
  \csname #3\endcsname}%
\else
\gdef\SetFigFont#1#2#3{\begingroup
  \count@#1\relax \ifnum 25<\count@\count@25\fi
  \def\x{\endgroup\@setsize\SetFigFont{#2pt}}%
  \expandafter\x
    \csname \romannumeral\the\count@ pt\expandafter\endcsname
    \csname @\romannumeral\the\count@ pt\endcsname
  \csname #3\endcsname}%
\fi
\fi\endgroup
\begin{picture}(4304,1456)(522,-2364)
\put(4550,-1861){\makebox(0,0)[lb]{\smash{\SetFigFont{10}{12.0}{rm}$V$}}}
\put(3376,-1561){\makebox(0,0)[lb]{\smash{\SetFigFont{10}{12.0}{rm}$+ \sum$}}}
\put(522,-1561){\makebox(0,0)[lb]{\smash{\SetFigFont{10}{12.0}{rm}$D_5(\kf_1,\kf_2,\kf_3,\kf_4,\kf_5) = \:\sum$}}}
\end{picture}

\ee
The summation in the last term now refers to the possibilities to 
combine two of the five momenta into a sum. 

This mechanism seems to apply to every 
odd number of intermediate gluons, 
and only even numbers of gluons appear as intermediate states.  
What we observe here is a generalization of the so--called reggeiziation 
of the gluon, a rather deep property of QCD in the Regge limit. 

The next step will be the calculation of 
the 6--gluon amplitude. From that we will learn more about 
the elements of the field theory. The first question to answer is 
whether there is a new $2\rightarrow 6$ transition vertex. 
Further, we will be able to understand the transition from the 4--gluon 
to the 6--gluon system. 
This should involve the known vertex $V_{2\rightarrow 4}$, but also 
its generalization to the case in which the 2 gluons above the 
vertex are not in a color singlet, as has been the case in the 4--gluon 
amplitude. 

\section{Conformal Invariance}
The Lipatov kernel and the transition vertex $V_{2\rightarrow 4}$ 
exhibit a high degree of symmetry in that they are conformally 
invariant. To explain the meaning of this conformal invariance 
we go by means of Fourier transformation from transverse momentum 
space (momenta $\{\kf\}$) to impact parameter 
space (vectors $\{\vec{\rho} \}$). We then introduce the 
complex coordinates $\rho = \rho_x +i \rho_y$ and 
$\rho^\ast = \rho_x - i \rho_y$. 
This is done for all arguments of the Lipatov kernel and of the 
vertex. 

M\"obius (or conformal) transformations are defined by 
\be
   \rho \rightarrow \rho^\prime = \frac{a \rho +b}{c\rho +d}\,;\:\:\:\:
   ad -bc = 1  \,.
\label{mob}
\ee
These transformations are characterized by the group 
\be
  \left(
  \begin{array}{cc}
  {a}&{b}\\
  {c}&{d}
  \end{array}
  \right)
  \in SL(2,{\bf{C}}) / Z_2 \,,
\ee
i.\,e.\ the group of projective conformal transformations. 
The generators of this group form the subalgebra $sl(2,{\bf C})$ of 
the well-known Virasoro algebra. 
It was known for some time\,\cite{BFKLconf} that the Lipatov kernel 
is invariant under the transformations (\ref{mob}). 
Recently, also the vertex $V_{2 \rightarrow 4}$ was shown to 
be symmetric under conformal transformations.\cite{Vconf}
We expect that this will be true also for higher vertices. 

The observation that the Lipatov kernel as well as the number 
changing vertex are conformally invariant naturally leads us to 
a further property of the effective theory: it will 
be a conformal field theory. 
Conformal field theories in two dimensions have been subject 
to very intense and fruitful investigation in the past years.\cite{CFT} 
The applications range from statistical physics to string theory. 
It turned out that conformal symmetry is an extremely powerful tool. 
One example which might possibly apply to our considerations 
is the fact that the $n$--point functions in a conformal field theory 
are highly restricted. The 3--point function, for instance, is 
fixed up to a constant. 
The connection with conformal field theory might be very useful 
for a deeper understanding of small--$x$ QCD. 

It is important to mention that the conformal symmetry described above 
is not an exact symmetry of Nature even in the Regge limit. 
It is known that the running of the gauge coupling 
$\alpha_s$ (the gauge coupling is fixed in LLA) 
and possibly other next--to--leading corrections will 
break conformal invariance. 
Of course, it will be important to understand in 
detail how the symmetry breaking takes place. 

\section{Outlook}
The unitarization of the BFKL pomeron is urgently needed 
to get a deeper understanding of 
the small--$x$ behavior of structure functions. 
It is believed that the unitarization of the pomeron will 
lead us to an effective theory of QCD in the Regge limit. 

We are still at the very beginning of constructing such an 
effective theory of small--$x$ QCD. We have compelling evidence that 
this theory will be a $2+1$ dimensional conformal field theory. 
Reggeized gluons have been identified as the correct degrees of 
freedom. Some of the elements of the effective theory are 
already known. Other very important elements are still missing, 
but there has been considerable progress in the last years. 
So far, the emerging picture is very encouraging. 

\section*{Acknowledgments}
I would like to thank J.\ Bartels, H.\ Lotter and M.\ W\"usthoff 
for many helpful discussions.

\end{document}